\documentclass[manuscript]{aastex}
\usepackage{natbib}
\usepackage{amsmath,amssymb}
\usepackage{empheq}
\usepackage[titletoc,title]{appendix}
\usepackage{epstopdf}

\begin{document}
	\title{Modeling of Joint Parker Solar Probe -- Metis/Solar Orbiter Observations }
	
	\author{L. Adhikari$^{1}$, G. P. Zank$^{1}$, D. Telloni$^2$, and L.-L. Zhao$^{1}$ }
	
	\altaffiltext{1}{Center for Space Plasma and Aeronomic Research (CSPAR), and Department of Space Science, University of Alabama in Huntsville, Huntsville, AL 35899, USA}
	
	 \altaffiltext{2}{National Institute for Astrophysics—Astrophysical Observatory of Torino Via Osservatorio 20, I-10025 Pino Torinese, Italy}
	
	
		
	


\begin{abstract}
We present a first theoretical modeling of joint Parker Solar Probe (PSP) - Metis/Solar Orbiter (SolO) quadrature observations \citep{Telloni2022}. The combined observations describe the evolution of a slow solar wind plasma parcel from the extended solar corona ($3.5-6.3$ R$_\odot$) to the very inner heliosphere (23.2 R$_\odot$). The Metis/SolO instrument remotely measures the solar wind speed finding a range from $96-201$ kms$^{-1}$, and PSP measures the solar wind plasma in situ, observing a radial speed of 219.34 kms$^{-1}$. We find theoretically and observationally that the solar wind speed accelerates rapidly within 3.3 -- 4 R$_\odot$, and then increases more gradually with distance. Similarly, we find that the theoretical solar wind density is consistent with the remotely and in situ observed solar wind density. The normalized cross-helicity and normalized residual energy observed by PSP are 0.96 and -0.07, respectively, indicating that the slow solar wind is very Alfv\'enic. The theoretical NI/slab results are very similar to PSP measurements, which is a consequence of the highly magnetic field-aligned radial flow ensuring that PSP can measure slab fluctuations and not 2D. Finally, we calculate the theoretical 2D and slab turbulence pressure, finding that the theoretical slab pressure is very similar to that observed by PSP.   	
\end{abstract}	
\keywords{ magnetohydrodynamics (MHD) -- plasmas -- turbulence -- Sun: corona -- Sun: heliosphere -- solar wind}
	
\section{Introduction} 
One of the main purposes of Parker Solar Probe (PSP) and Solar Orbiter (SolO) is to understand coronal heating, and the acceleration of the solar wind. Turbulence is thought to be a central element in addressing these questions. PSP and SolO allow us to study the radial evolution of turbulence in the inner heliosphere \citep[e.g.,][]{2021ApJ...920L..14T,A2022,Telloni2022}. \cite{A2022} studied the evolution of 2D and slab turbulence in similar types of slow solar wind observed by PSP \citep{2016SSRv..204..131K,2016SSRv..204...49B} and SolO \citep{2020A&A...642A...9H,2020A&A...642A..16O}, finding that 2D turbulence is the dominant component. \cite{2021ApJ...920L..14T} studied the evolution of turbulence in the same plasma parcel measured by PSP and SolO, using a radial alignment between PSP and SolO from 0.1 au (PSP's position) to 1 au (SolO's position). This alignment facilitates the study of turbulence characteristics on the same plasma parcel with different locations.  

When PSP was at 0.11 au on January 18, 2021 at 18:59 UT during encounter 7, PSP entered the plane of sky (POS) measured by the SolO coronagraph Metis on January 17, 2021 at 16:30 UT \cite[][interval \# 1]{Telloni2022}. Five days later, when PSP was at 0.26 au on January 23, 2021 at 17:02 UT, PSP again crossed the POS corresponding to the longitude observed by Metis \cite[][interval \# 2]{Telloni2022}.
\cite{Telloni2022} presented the first observational study that combined in situ PSP data and remote Metis/SolO observations. Their unique configuration allowed Telloni et al to follow the evolution of a plasma parcel through the extended solar corona ($3.5-6.3$ R$_\odot$, where R$_\odot=6.95 \times 10^5$ km) to the very inner heliosphere. \cite{Telloni2022} therefore provide plasma observations from the sub-Alfv\'enic corona to the super-Alfv\'enic solar wind.      

Using a Potential Field Source Surface (PFSS) extrapolation \citep[e.g.,][]{2020ApJS..246...54P},
\cite{Telloni2022} found that PSP sampled (interval \# 1) plasma coming from the equatorial extension of the southern polar coronal hole, and (interval \# 2) plasma coming from the low latitude of a northern coronal hole. During interval \# 1, PSP observed almost exclusively outwardly propagating Alfv\'en waves ($\sigma_c \sim 1$, where $\sigma_c$ is the normalized cross-helicity), whereas interval \# 2 contains switchbacks \citep{2019Natur.576..237B,Telloni2022}. In this letter, we model the solar wind plasma (interval \# 1) measured by the PSP -- SolO quadrature from the sub- to super-Alfv\'enic solar wind, and compare the predicted solar wind radial profile with the combined PSP (in situ) -- Metis (remote data set). Our model describes the (coronal) heating of the slow solar wind near the equatorial region and the acceleration of the solar wind. This differs from heating and acceleration in open field region \citep{1999ApJ...523L..93M,2001ApJ...548..482D,2009ApJ...707.1659C,2013ApJ...767..125C,2010ApJ...708L.116V,2018ApJ...854...32Z,2010ApJ...720..824C,2020ApJ...901..102A,2022ApJ...929...98T}, although the heating mechanism is thought to be the same for the fast and slow solar wind flow \citep{2021PhPl...28h0501Z}. 

We structure the letter as follows. Section 2 discusses a solar wind model that incorporates nearly incompressible magnetohydrodynamic (NI MHD) turbulence. Section 3 discusses the data analysis. Section 4 compares the theoretical and observed results. Finally, Section 5 is the conclusions.
\section{A turbulent solar wind model}
In the letter, we use a superradial expansion turbulent driven solar model \citep{Telloni2022b} to study the coronal plasma in the slow solar wind that the hot plasma emerges from the closed loop into the open field region by interchange reconnection, after which it expands superradially.
We consider a steady flow in a one-dimensional, superradially expanding open flux tube of cross-sectional area $A(r) (=r^2 f(r)$, $f(r)$ is a superradial expansion factor), and $A(r)$ is inversely proportional to the magnetic field strength $B(r)$, 
\begin{equation}
B_r A(r) =  B_r r^2 f(r) = const.
\end{equation}
The super radial expansion term $f(r)$ is given by \citep{1976SoPh...49...43K},
\begin{equation}
f(r) = \frac{f_m \exp(\frac{r-r_a}{\sigma}) + 1 - (f_m-1)\exp(\frac{R_\odot-r_a}{\sigma}) }{\exp (\frac{r-r_a}{\sigma}) + 1},
\end{equation}
where $f_m=2$, $r_a=2$ R$_\odot$, and $\sigma = 0.8$ R$_\odot$.
The steady flow in the superradially expanding tube can be described by the continuity, inviscid momentum, and pressure equations,
\begin{equation}
\frac{d n_s}{dr}=- \frac{2n_s}{r} - \frac{n_s}{U} \frac{dU}{dr} - \frac{n_s}{\sigma f(r)} \exp\bigg(\frac{r-r_a}{\sigma} \bigg) \frac{f_m - f(r)}{\exp(\frac{r-r_a}{\sigma}) + 1}  ;
\end{equation}
\begin{equation}
\rho U \frac{dU}{dr} = - \frac {d P}{dr} - \frac{GM_\odot}{r^2} \rho;
\end{equation}
\begin{equation}
\frac{dP}{dr} = - \frac{\gamma P}{U} \frac{dU}{dr} - \frac{2\gamma P}{r}  -  \frac{\gamma P}{\sigma f(r)} \exp\bigg(\frac{r-r_a}{\sigma} \bigg) \frac{f_m - f(r)}{\exp(\frac{r-r_a}{\sigma}) + 1} + (\gamma -1) s_1 \frac{S_t}{U},
\end{equation}
where $n_s(\rho)$ is the solar wind (mass) density, $U$ the solar wind speed, $P$ the thermal pressure, $G$ the gravitational constant, $M_\odot$ the solar mass, $S_t$ the turbulent heating term, and $\gamma(=5/3)$ the polytropic index. The parameter $s_1$ denotes the fraction of turbulence energy used to heat the coronal plasma (protons). We use $s_1=0.6$, meaning that the 60\% of the turbulent energy heats the coronal/solar wind plasma \citep{2009ApJ...702.1604C,2009JGRA..114.9103B,2018ApJ...856..159E,2019ApJS..242...12C,2021A&A...650A..16A,Telloni2022b}. 
Combining the above equations yields 
\begin{equation}
\frac{C_s^2}{U^2} \bigg(M_s^2 - 1 \bigg) \frac{dU}{dr} = \frac{2 \gamma P}{\rho U r} - (\gamma - 1) s_1 \alpha \frac{S_t}{m_p n_s U^2} - \frac{GM_\odot}{Ur^2} + \frac{\gamma P}{\sigma f(r) \rho U} \exp\bigg(\frac{r-r_a}{\sigma} \bigg) \frac{f_m - f(r)}{\exp(\frac{r-r_a}{\sigma}) + 1},
\end{equation}
where $M_s = U/C_s$ is the sonic Mach number, and $C_s^2 = \gamma P/\rho$ is the square of the sound speed. Equation (6) possesses a critical point, where $M_s=1$ and the right hand side (rhs) is zero simultaneously. We use L'H$\hat{\text{o}}$pital's rule to solve Equation (6) in the vicinity of the critical point. This solar wind model (Equations (3) -- (5)) includes only the thermal force, and not the ponderomotive force \citep[see,][]{1970ARA&A...8...31H,1982SSRv...33..161L,1988ApJ...325..442W,1999JGR...10419765F,2013ApJ...767..125C,2010ApJ...708L.116V,2010ApJ...720..824C} or wave pressure \citep{1995A&A...303L..45M}. The turbulent heating term $S_{t}$ can be derived from a von K\'arm\'an phenomenology, and is given by \citep{2010ApJ...708L.116V,2015ApJ...805...63A,2018ApJ...869...23Z} 
\begin{equation}
\begin{split}  
S_{t} & = \alpha m_{p} n_{s} \bigg[ 2 \frac{\langle {{z^\infty}^+}^2 \rangle^2 \langle {{z^\infty}^-}^2 \rangle^{1/2}}{L_\infty^+} + 2 \frac{\langle {{z^\infty}^-}^2 \rangle^2 \langle {{z^\infty}^+}^2 \rangle^{1/2}}{L_\infty^-} + 2 \frac{\langle {{z^*}^+}^2 \rangle \langle  {{z^\infty}^+}^2  \rangle \langle  {{z^\infty}^-}^2  \rangle^{1/2}}{L_\infty^+}   \bigg],
\end{split} 
\end{equation}
where $\langle {{z^\infty}^\pm}^2 \rangle$ are the 2D outward and inward Els\"asser energies, $L^\pm_\infty$ the corresponding energy weighted correlation lengths, and $\langle {{z^*}^+}^2 \rangle$ is the NI/slab energy in forward propagating modes. The parameter $\alpha(=0.01)$ is a von K\'arm\'an-Taylor constant, and $m_p$ the proton mass. 

The 1D steady-state transport equations for the majority 2D turbulence, including the superradial expansion factor are given by \citep{2017ApJ...835..147Z,2020ApJ...901..102A,Telloni2022b}
\begin{equation}
\begin{split}  
U \frac{d \langle {z^\infty}^{\pm 2} \rangle}{d r} & = - \bigg(\frac{\langle {z^\infty}^{\pm 2} \rangle}{2} + \bigg(2a - \frac{1}{2} \bigg) E_D^\infty \bigg) \frac{dU}{dr} - \frac{2 U}{r} \bigg(\frac{\langle {z^\infty}^{\pm 2} \rangle}{2} + \bigg(2a - \frac{1}{2} \bigg) E_D^\infty \bigg)  - \frac{U}{\sigma f(r)} \bigg(\frac{\langle {z^\infty}^{\pm 2} \rangle}{2} \\
& + \bigg(2a - \frac{1}{2} \bigg) E_D^\infty \bigg) \exp\bigg(\frac{r-r_a}{\sigma} \bigg) \frac{f_m - f(r)}{\exp(\frac{r-r_a}{\sigma}) + 1}  - 2 \alpha \frac{\langle {z^\infty}^{\pm 2} \rangle^2 \langle {z^\infty}^{\mp 2} \rangle^{1/2}  }{L^\pm_\infty} + S^{\langle {z^\infty}^{\pm 2} \rangle};
\end{split} 
\end{equation}	
\begin{equation}
\begin{split}  
U \frac{d E_D^\infty}{d r} & = - \bigg(\frac{E_D^\infty}{2} + \bigg(2a - \frac{1}{2} \bigg) E_T^\infty \bigg) \frac{dU}{dr} - \frac{2 U}{r} \bigg(\frac{E_D^\infty}{2} + \bigg(2a - \frac{1}{2} \bigg) E_T^\infty \bigg)  - \frac{U}{\sigma f(r)} \bigg(\frac{E_D^\infty}{2} \\
& + \bigg(2a - \frac{1}{2} \bigg) E_T^\infty \bigg) \exp\bigg(\frac{r-r_a}{\sigma} \bigg) \frac{f_m - f(r)}{\exp(\frac{r-r_a}{\sigma}) + 1}  - \alpha E_D^\infty \bigg(\frac{\langle {z^\infty}^{+ 2} \rangle^{1/2} \langle {z^\infty}^{- 2} \rangle }{L_\infty^-} + \frac{\langle {z^\infty}^{- 2} \rangle^{1/2} \langle {z^\infty}^{+ 2} \rangle  }{L_\infty^+}   \bigg) \\
& + S^{E_D^\infty};
\end{split} 
\end{equation}	
\begin{equation}
\begin{split}  
U \frac{dL_\infty^\pm}{dr} & = - \bigg(\frac{L_\infty^\pm}{2} + \bigg(a - \frac{1}{4} \bigg) L_D^\infty\bigg) \frac{dU}{dr} - \frac{2U}{r} \bigg(\frac{L_\infty^\pm} {2} + \bigg(a - \frac{1}{4} \bigg) L_D^\infty \bigg )  - \frac{U}{\sigma f(r)} \bigg(\frac{L_\infty^\pm} {2} + \bigg(a - \frac{1}{4} \bigg) L_D^\infty \bigg ) \\
& \times \exp\bigg(\frac{r-r_a}{\sigma} \bigg) \frac{f_m - f(r)}{\exp(\frac{r-r_a}{\sigma}) + 1};
\end{split}  
\end{equation}
\begin{equation}
\begin{split}  
U \frac{dL_D^\infty}{dr} & = - \bigg(\frac{L_D^\infty}{2} + \bigg(2a - \frac{1}{2} \bigg) (L_\infty^+ + L_\infty^-) \bigg) \frac{dU}{dr} - \frac{2U}{r} \bigg(\frac{L_D^\infty} {2} + \bigg(2a - \frac{1}{2} \bigg) (L_\infty^+ + L_\infty^-) \bigg )  - \frac{U}{\sigma f(r)} \\
& \times \bigg(\frac{L_D^\infty} {2} + \bigg(2a - \frac{1}{2} \bigg) (L_\infty^+ + L_\infty^-) \bigg ) \exp\bigg(\frac{r-r_a}{\sigma} \bigg) \frac{f_m - f(r)}{\exp(\frac{r-r_a}{\sigma}) + 1},
\end{split}  
\end{equation}
where $L_D^\infty$ is the energy weighted correlation length for the 2D residual energy $E_D^\infty$, and $E_T^\infty$ is the 2D total turbulence energy. The term \textquotedblleft S\textquotedblright~refers to the turbulent shear source for the 2D outward and inward Els\"asser energies, and the residual energy. Equation (8) can be written in terms of $E_T^\infty=(\langle z^{\infty + 2} \rangle + \langle z^{\infty - 2} \rangle)/2 $, and the cross-helicity $E_C^\infty=(\langle z^{\infty + 2} \rangle - \langle z^{\infty - 2} \rangle)/2 $ as
\begin{equation}
\begin{split}
& U \frac{d E_T^\infty}{dr} = -\bigg(\frac{E_T^\infty}{2} + \bigg(2a - \frac{1}{2} \bigg) E_D^\infty  \bigg) \frac{dU}{dr} - \frac{2U}{r} \bigg(\frac{E_T^\infty}{2} + \bigg(2a - \frac{1}{2} \bigg) E_D^\infty  \bigg)  - \frac{U }{\sigma f(r)} \\
& \times \bigg(\frac{E_T^\infty}{2} + \bigg(2a - \frac{1}{2} \bigg) E_D^\infty  \bigg)\exp\bigg(\frac{r-r_a}{\sigma}\bigg) \frac{f_m - f(r)}{\exp\big(\frac{r-r_a}{\sigma} \big)+1} -  \alpha \frac{|E_T^\infty + E_C^\infty |^2 |E_T^\infty - E_C^\infty|^{1/2} }{L_\infty^+} \\
& -  \alpha \frac{|E_T^\infty - E_C^\infty |^2 |E_T^\infty + E_C^\infty|^{1/2} }{L_\infty^+} + \frac{S^{\langle z^{\infty + 2} \rangle} +  S^{\langle z^{\infty - 2} \rangle}}{2};
\end{split}
\end{equation}
\begin{equation}
\begin{split}
& U\frac{dE_C^\infty}{dr} = - \frac{E_C^\infty}{2} \frac{dU}{dr} - \frac{U}{r} E_C^\infty - \frac{U}{2\sigma f(r)} E_C^\infty \exp\bigg(\frac{r-r_a}{\sigma}\bigg) \frac{f_m - f(r)}{\exp\big(\frac{r-r_a}{\sigma} \big)+1} \\
& -  \alpha \frac{|E_T^\infty + E_C^\infty |^2 |E_T^\infty - E_C^\infty|^{1/2} }{L_\infty^+}  +  \alpha \frac{|E_T^\infty - E_C^\infty |^2 |E_T^\infty + E_C^\infty|^{1/2} }{L_\infty^+} + \frac{S^{\langle z^{\infty + 2} \rangle} -  S^{\langle z^{\infty - 2} \rangle}}{2}.
\end{split}
\end{equation}
Equation (12) can be written in the conservation form \citep{2022ApJ...928..176W},
\begin{equation}
\begin{split} 
& \frac{1}{r^2 f(r)} \frac{d}{dr} \bigg[r^2 f(r) U (E^\infty_w + P^\infty_w) \bigg] = U \frac{dP_w^\infty}{dr} +   \rho \bigg[- \alpha \frac{|E_T^\infty + E_C^\infty |^2 |E_T^\infty - E_C^\infty|^{1/2} }{L_\infty^+}  \\
& - \alpha \frac{|E_T^\infty - E_C^\infty |^2 |E_T^\infty + E_C^\infty|^{1/2} }{L_\infty^+} + \frac{S^{\langle z^{\infty + 2} \rangle} +  S^{\langle z^{\infty - 2} \rangle}}{2}  \bigg],
\end{split} 
\end{equation}
where $E_w^\infty$ = $\rho E_T^\infty/2$ is the 2D turbulence energy density, and 
\begin{equation}
P_w^\infty = \frac{\rho}{2} \bigg[ \frac{E_T^\infty}{2} + \bigg(2a - \frac{1}{2} \bigg) E_D^\infty \bigg]  = \frac{E_w^\infty}{2} \bigg[1 + 2 \bigg(2a - \frac{1}{2} \bigg) \sigma_D^\infty   \bigg]
\end{equation}
is the 2D turbulence pressure. In the absence of the nonlinear term and turbulence source terms, Equation (14) resemble the well-known Wentze-Kramers-Brillouin (WKB) form. The terms in the square bracket on the left hand side (lhs) of Equation (14) express the energy density flux vector, which describes the amount of turbulence energy per unit time per unit area in a direction perpendicular to the velocity \citep{1987flme.book.....L}. Similarly, on the lhs, the first term in [...] describes the energy transmitted through the unit surface area per unit time, and the second term describes the work done by the turbulence pressure on the plasma in the surface. The first term on the right hand side describes the rate of turbulence pressure gradient on the background plasma flow.

The 1D steady-state transport equations for the energy in NI/slab forward propagating modes and the corresponding energy weighted correlation length are \citep{2017ApJ...835..147Z,2020ApJ...901..102A,Telloni2022b},
\begin{equation}
\begin{split} 
(U - V_A) \frac{d \langle {z^*}^{+ 2} \rangle}{dr} & = -\frac{1}{2} \frac{dU}{dr}\langle {z^*}^{+ 2} \rangle + (2b-1) \frac{U}{r} \langle {z^*}^{+ 2} \rangle_+ \frac{V_A}{2 \rho} \frac{d\rho}{dr} \langle {z^*}^{+ 2} \rangle  + \frac{1}{2} (2b-1) \frac{U \langle {z^*}^{+ 2} \rangle} {\sigma f(r)} \\
& \times \exp\bigg(\frac{r-r_a}{\sigma} \bigg) \frac{f_m - f(r)}{\exp(\frac{r-r_a}{\sigma}) + 1}  - 2 \alpha \frac{\langle {z^*}^{+ 2} \rangle \langle {z^\infty}^{- 2} \rangle^{1/2}}{\lambda_\infty^+} + S^{\langle {z^*}^{+ 2} \rangle};
\end{split}  
\end{equation}
\begin{equation}
\begin{split}
(U - V_A) \frac{dL_*^+}{dr} & = - \frac{1}{2} \frac{dU}{dr} L_*^+ + (2b-1) \frac{U}{r} L_*^+  + \frac{V_A}{2 \rho} \frac{d\rho}{dr} L_*^+  + \frac{1}{2} (2b-1) \frac{U L_*^+}{\sigma f(r)} \exp\bigg(\frac{r-r_a}{\sigma} \bigg) \frac{f_m - f(r)}{\exp(\frac{r-r_a}{\sigma}) + 1}.
\end{split}
\end{equation}
The parameter $V_A(= ({B_0}/{\sqrt{\mu_0 \rho}}) ({r_0}/{r} )^2 (1/f(r))$, where $B_0$ is the magnetic field at a reference point $r_0$, and $\mu_0$ is the magnetic permeability) is the large-scale Alfv\'en velocity. The parameter $S^{\langle {z^*}^{+ 2} \rangle}$ denotes the turbulent shear source for the energy in NI/slab forward propagating modes. We use $b=0.26$ \citep[see][for further discussion]{2012ApJ...745...35Z,2017ApJ...835..147Z}. Equation (16) can be written in terms of the slab total turbulent energy $E_T^* (= \langle z^{* + 2} \rangle/2 \equiv E_C^*$, $\langle z^{* - 2} \rangle=0$) as
\begin{equation}
\begin{split}
& (U-V_A) \frac{dE_T^*}{dr} = -\frac{1}{2} \frac{dU}{dr} E_T^* + (2b-1) \frac{U}{r} E_T^* + \frac{V_A}{2 \rho} \frac{d\rho}{dr} E_T^* + \frac{1}{2} (2b-1) \frac{U}{\sigma f(r)} E_T^* \\
& \times \exp\bigg(\frac{r-r_a}{\sigma} \bigg) \frac{f_m - f(r)}{\exp(\frac{r-r_a}{\sigma}) + 1} - 2 \alpha \frac{E_T^* |E_T^\infty + E_C^\infty||E_T^\infty - E_C^\infty|^{1/2}}{L_\infty^+}  + \frac{S^{\langle z^{* + 2} \rangle}}{2}.
\end{split}
\end{equation}
Equation (18) can also be written in the conservation form \citep{2022ApJ...928..176W},
\begin{equation}
\begin{split}
& \frac{1}{r^2 f(r)} \frac{d}{dr} \bigg[r^2 f(r) \big((U - V_A) E_w^*   + U P_w^*\big)  \bigg] = U \frac{dP_w^*}{dr} + 2E_w^* \bigg(4b\frac{u}{r} + 2b \frac{U}{f(r)}\frac{df(r)}{dr} \bigg) \\
& + \frac{\rho}{2} \bigg[ - 2 \alpha \frac{E_T^* |E_T^\infty + E_C^\infty||E_T^\infty - E_C^\infty|^{1/2}}{L_\infty^+}  + \frac{S^{\langle z^{* + 2} \rangle}}{2}  \bigg],
\end{split}
\end{equation}
where $E_w^*=\rho E_T^*/2$ is the slab turbulence energy density, and $P_w^*=E_w^*/2$ is the slab turbulence pressure. Equation (19) also resembles the WKR form in the absence of the mixing term, dissipation terms, and the turbulence source term.

Equations (8)--(13) and (16) -- (18) are a set of turbulence transport equations describing the evolution of turbulence in the highly field-aligned flows \cite[see][for a detailed discussion]{2020ApJ...901..102A}.

Similar to \cite{Telloni2022b}, we use two forms of the turbulent shear source, i) in region between the sonic surface and Alfv\'en surface, where the sound speed $C_S$ is assumed to be the characteristic speed, and ii) beyond the Alfv\'en surface, where the Alfv\'en speed is considered to be the characteristic speed.
The shear source of turbulence in the region between the sonic and Alfv\'en surfaces can be written in the form \citep{Telloni2022b}
\begin{equation}
\begin{split} 
S^s_{\langle {z^{\infty,*}}^{\pm 2} \rangle} = C_{\infty,*}^{s+} \frac{r_0^s |\Delta U| C^2_{s}}{r^2}=\gamma C_{\infty,*}^{s+}\frac{r_0^s |U - U_0^{s'}| P}{\rho r^2}; \\\quad S^s_{\langle {z^{\infty}}^{- 2} \rangle} = C_{\infty}^{s-} \frac{r_0^s |\Delta U| C^2_{s}}{r^2} = \gamma C_{\infty}^{s-} \frac{r_0^s |U - U_0^{s'}| P}{\rho r^2}; \\
\quad S^{s}_{E_D^\infty}  = C_{\infty}^{s E_D} \frac{r_0^s |\Delta U| C^2_{s}}{r^2}=\gamma C_{\infty}^{s E_D} \frac{r_0^s |U - U_0^{s'}| P}{\rho r^2},
\end{split} 
\end{equation} 
where $U_0^{s'}$ is the solar wind speed at $r_0^s$, a position above the sonic surface, and $C^{s \pm,E_D}_{\infty,*}$ denotes the strength of the shear source of turbulence. Here we use $\Delta U = |U - U_0^{s'}|$, and $C_s^2=\gamma P/\rho$, which results in the shear source of turbulence depending on distance $r$, solar wind speed $U$, the thermal pressure $P$, and the proton mass density $\rho$. We use $r_0^s=5.16$ R$_\odot$, and $U_0^{s'}=173.5$ kms$^{-1}$.

Similarly, the turbulent shear source above the Alfv\'en surface is,
\begin{equation}
\begin{split}  
S^A_{\langle {z^{\infty,*}}^{\pm 2} \rangle} = C_{\infty,*}^{a+} \frac{r_0^a |\Delta U| V^2_{A}}{r^2} = C_{\infty,*}^{a+} \frac{r_0^a |U-U_0^{a'}| V^2_{A}}{r^2}; \\ \quad S^A_{\langle {z^{\infty}}^{- 2} \rangle} = C_{\infty}^{a-} \frac{r_0^a |\Delta U| V^2_{A0}}{r^2}=C_{\infty}^{a-}\frac{r_0^a |U-U_0^{a'}| V^2_{A}}{r^2};\\ \quad S^{A}_{E_D^\infty}  = C_{\infty}^{a E_D} \frac{r_0^a |\Delta U| V^2_{A0}}{r^2}=C_{\infty}^{a E_D}\frac{r_0^a |U-U_0^{a'}| V^2_{A}}{r^2},
\end{split} 
\end{equation} 
where we use $\Delta U = |U - U_0^{a'}|$. The parameter $U_0^{a'}$ is the solar wind speed at $r_0^a$, a position above the Alfv\'en surface, and $C^{a \pm,E_D}_{\infty,*}$ denote the strength of the shear source of turbulence. We use $r_0^a=9.22$ R$_\odot$, and $U_0^{a'}=222.87$ kms$^{-1}$.

\section{Data analysis}	
We calculate the transverse turbulence energy and the transverse correlation length using a method developed by \cite{A2022}. A fluctuating vector ${\bf a} (= a_r \hat{r} + a_t \hat{t} + a_n \hat{n}$, where $a_r$, $a_t$, $a_n$ are the $R$, $T$, and $N$ components of a vector ${\bf a}$) can be decomposed into parallel and perpendicular vectors relative to the mean magnetic field ${\bf B}$ as
\begin{equation*}
{\bf a} = {\bf a}_{||} + {\bf a}_\perp = a_{||} \hat{b} + {\bf a}_\perp,
\end{equation*} 
where $\hat{b}={\bf B}/|{\bf B}|$ is the unit vector, $|{\bf B}|$ is the magnitude of the mean magnetic field ${\bf B} (= B_R \hat{r} + B_T \hat{t} + B_N \hat{n}$, where $B_R$, $B_T$, and $B_N$ denote the $R$, $T$, and $N$ components of the mean magnetic field), and ${\bf a}_\perp$ and ${\bf a}_{||}(=a_{||} \hat{b})$ are the perpendicular and parallel vectors, respectively. The parallel component $a_{||}$ can be written as
\begin{equation}
a_{||} = {\bf a} \cdot \hat{b},
\end{equation}
and the perpendicular vector ${\bf a_\perp}$,
\begin{equation}
{\bf a}_\perp = - \hat{b} \times (\hat{b} \times {\bf a}) = \frac{({\bf B} \times {\bf a} ) \times {\bf B}}{|{\bf B}|^2},
\end{equation}
can be derived as \citep{A2022},
\begin{equation}
\begin{split}       
{\bf a}_\perp & = \frac{C_T B_N - C_N B_T }{B_R^2 + B_T^2 + B_N^2} \hat{r}  + \frac{C_N B_R - C_R B_N}{B_R^2 + B_T^2 + B_N^2} \hat{t}  + \frac{C_R B_T - C_T B_R}{B_R^2 + B_T^2 + B_N^2} \hat{n},
\end{split}  
\end{equation}
where $C_R = B_T A_N - B_N A_T$, $C_T = B_N A_R - B_R A_N$, and $C_N = B_R A_T - B_T A_R$. Using Equation (18), we calculate the transverse Els\"asser energies, fluctuating magnetic and kinetic energies, normalized cross-helicity and residual energy, and the corresponding transverse correlation lengths.

\section{Results}
In this section, we compare the theoretical and observed results from the extended solar corona to the very inner heliosphere.
We select SPAN ion plasma data \citep{2016SSRv..204..131K}, and 1 minute resolution FIELDS data \citep{2016SSRv..204...49B} in a time interval 18:40 -- 20:40 UT on January 2021 (interval \#1) during E7. We apply a boxcar method to the SPAN ion data to remove large spikes, and smooth the data using 1 minute long intervals, and then merge the plasma data with the magnetometer data. We calculate the transverse turbulence energy and the transverse correlation length, and the solar wind parameters for interval \# 1. The Metis/SolO plasma data (solar wind speed and density) are obtained from \cite{Telloni2022}.
We use Runge-kutta fourth order method to solve the solar wind + NI MHD turbulence transport equations from 3.3 -- 30 R$_\odot$. Table 1 shows the boundary conditions for the turbulence quantities and the solar wind parameters at 3.3 R$_\odot$. Table 2 shows the values of the strength of the turbulent shear source. These boundary conditions are chosen so the theoretical results are close to the observations. We note that the theoretical results are similar with or without the turbulent shear source. In this letter, we include the turbulent shear source, and compare the theoretical results with the observed results.
   
\begin{table}[h!]
	\centering
	\begin{tabular}{c c }
		\hline
		Parameters & Values \\ 
		\hline
		$\langle{{z^\infty}^\pm}^2\rangle$ (km$^2$s$^{-2}$) & $10^5$    \\
		$E_D^\infty$ (km$^2$s$^{-2}$) & 2000  \\
		$L_\infty^\pm$ (km$^3$s$^{-2}$) & $ 3.03 \times 10^{9}$   \\
		$L_D^\infty$ (km$^3$s$^{-2}$) & $1.6 \times 10^8$   \\
		$\langle{{z^*}^+}^2\rangle$ (km$^2$s$^{-2}$) & 6000    \\
		$L_*^+$ (km$^3$s$^{-2}$) & $9.1\times 10^7$ \\
		$U$ (kms$^{-1}$) & 45.13  \\
		$n$ (cm$^{-3}$) & $8 \times 10^{5}$   \\
		$T$ (K) & $7 \times 10^5$  \\
		\hline
	\end{tabular}
	\caption{Boundary values at 3.3 R$_\odot$ for the turbulent quantities and the solar wind parameters.  }
\end{table}

\begin{table}[h!]
	\centering
	\begin{tabular}{c c  }
		\hline
		Parameters & Values  \\ 
		\hline
		$C_\infty^{s \pm}$ & 0.1   \\
		$C_\infty^{s E_D}$ & -0.1   \\
		$C_*^{s+}$ & 0.1   \\
		$C_\infty^{a \pm}$ & 0.1  \\
		$C_\infty^{a E_D}$ & -0.1  \\
		$C_*^{a+}$ & 0.1   \\
		\hline
	\end{tabular}
	\caption{Assumed strengths for the shear driven quasi-2D and slab turbulence in the region between the sonic and Alfv\'en surfaces, and above the Alfv\'en surface.}
\end{table}	

\begin{figure}[h!]
	\centering
	\includegraphics[height=0.32\textwidth]{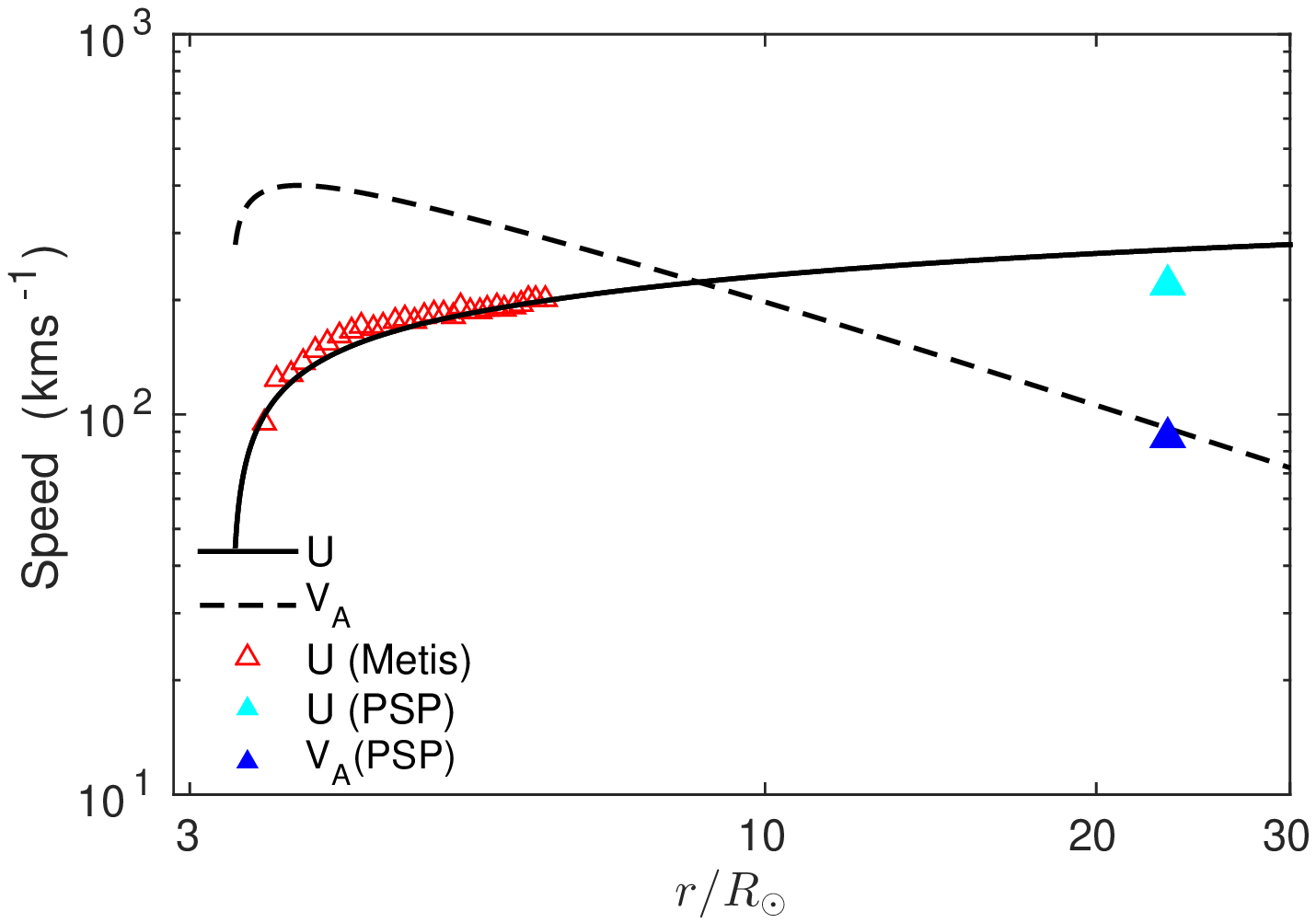}
    \includegraphics[height=0.32\textwidth]{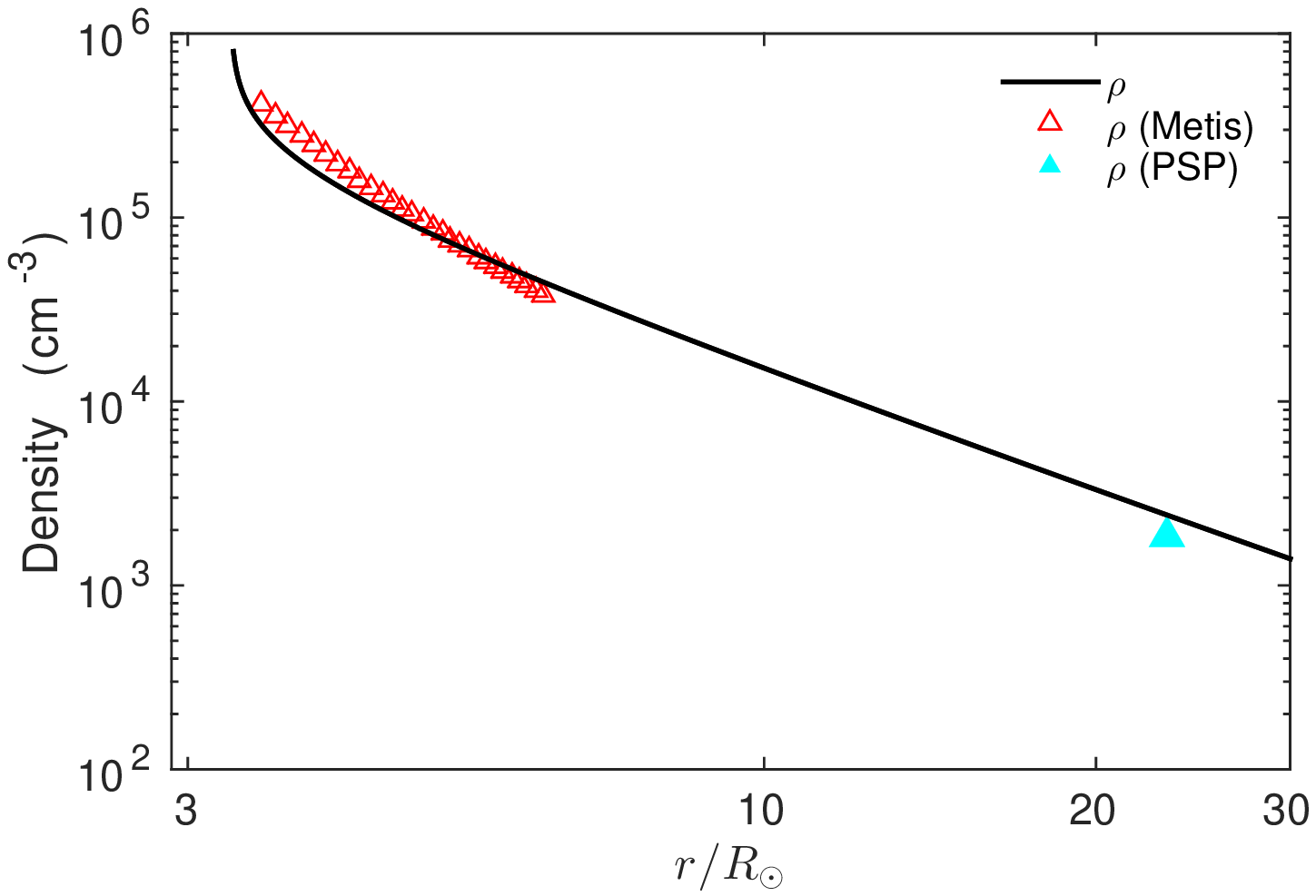}
	\caption{Comparison between the theoretical solar wind speed and Alfv\'en velocity (left), and the solar wind density (right) with the observed results of the plasma parcel measured by Metis/SolO \citep[obtained from][]{Telloni2022} and PSP. The solid and dashed curves are the theoretical results. The open red triangles are the observed speed and density from SolO/Metis, and the cyan full triangles are the observed speed and density, and the blue full triangle is the observed Alfv\'en speed measured by PSP during E7.  }
\end{figure}
During the PSP -- SolO Eastern-limb quadrature of mid-January, the same solar wind plasma stream was observed simultaneously in both the extended corona and the very inner heliosphere. The Metis coronagraph \citep{2020A&A...642A..10A} on board SolO imaged the $3.5$ to $6.3$ R$_{\odot}$ altitude range remotely, which corresponds to the coronal source region of the plasma flow impinging on PSP at $0.11$ au \citep[see][for a detailed description of the orbital geometry and connectivity during quadrature]{2021ApJ...920L..14T,Telloni2022}. The Metis instrument is designed to observe the solar corona both in polarized brightness (pB) and \ion{H}{1} Ly$\alpha$ ultraviolet (UV) emission. This allows for the study of complex coronal dynamics and structures. pB measurements are used to infer the electron density by exploiting the inversion technique developed by \citet{1950BAN....11..150V}. The UV light emitted by neutral hydrogen atoms is used as a proxy to estimate the outflow velocity of the proton component of the solar wind. The \ion{H}{1} Ly$\alpha$ line emission at $121.6$ nm is mainly due to resonant scattering processes of chromospheric radiation by coronal hydrogen atoms. It follows that an outward motion of the coronal plasma causes a reduction in intensity of the scattered Ly$\alpha$ line, since the incident radiation profile appears to be Doppler-shifted in the rest frame of the scattering atoms. This effect, known as Doppler dimming \citep{1987ApJ...315..706N}, can therefore be used to infer the expansion velocity of the coronal plasma. Based on a three-dimensional model of the large-scale solar corona (involving prior knowledge of electron density and temperature, kinetic temperature of scattering atoms, helium abundance, and temperature anisotropy, among others), the speed of the coronal flows is estimated by comparing observed and synthesized \ion{H}{1} Ly$\alpha$ line intensity \citep[see \S{} $11.2$ in][for an exhaustive review of the Doppler dimming technique]{2020A&A...642A..10A}. Figure 1 displays the modeled plasma $U$ and Alfv\'en $V_{A}$ speed, and proton number density, as a function of the altitude above the Sun. These profiles are compared with observations from SolO/Metis (open red triangles) and PSP (full triangles). Note that, accounting for a fully ionized plasma with $2.5\%$ helium \citep[according to][]{2020NatAs...4.1134M}, the electron density estimates provided by Metis were multiplied by $0.95$ to obtain the corresponding proton number density. The agreement between theory and observations is striking. The Metis and PSP measurements obviously are unrelated. One is obtained remotely and the other measured locally in situ. Evidently the model presented above reproduces very well the joint observations of the extended corona and the very inner heliosphere and the dynamic evolution of the solar wind plasma.

\begin{figure}[h!]
	\centering
	\includegraphics[height=0.32\textwidth]{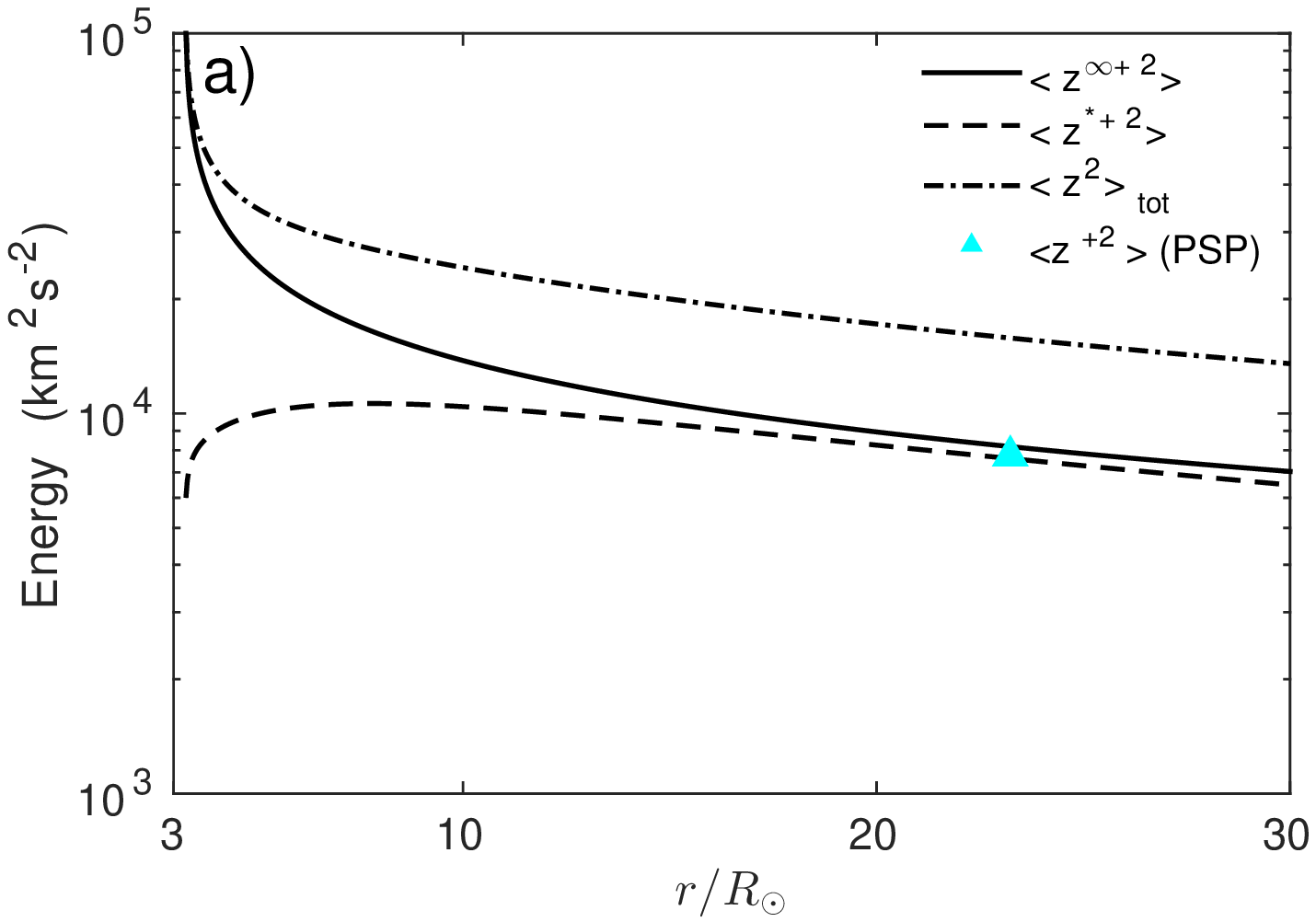}
	\includegraphics[height=0.32\textwidth]{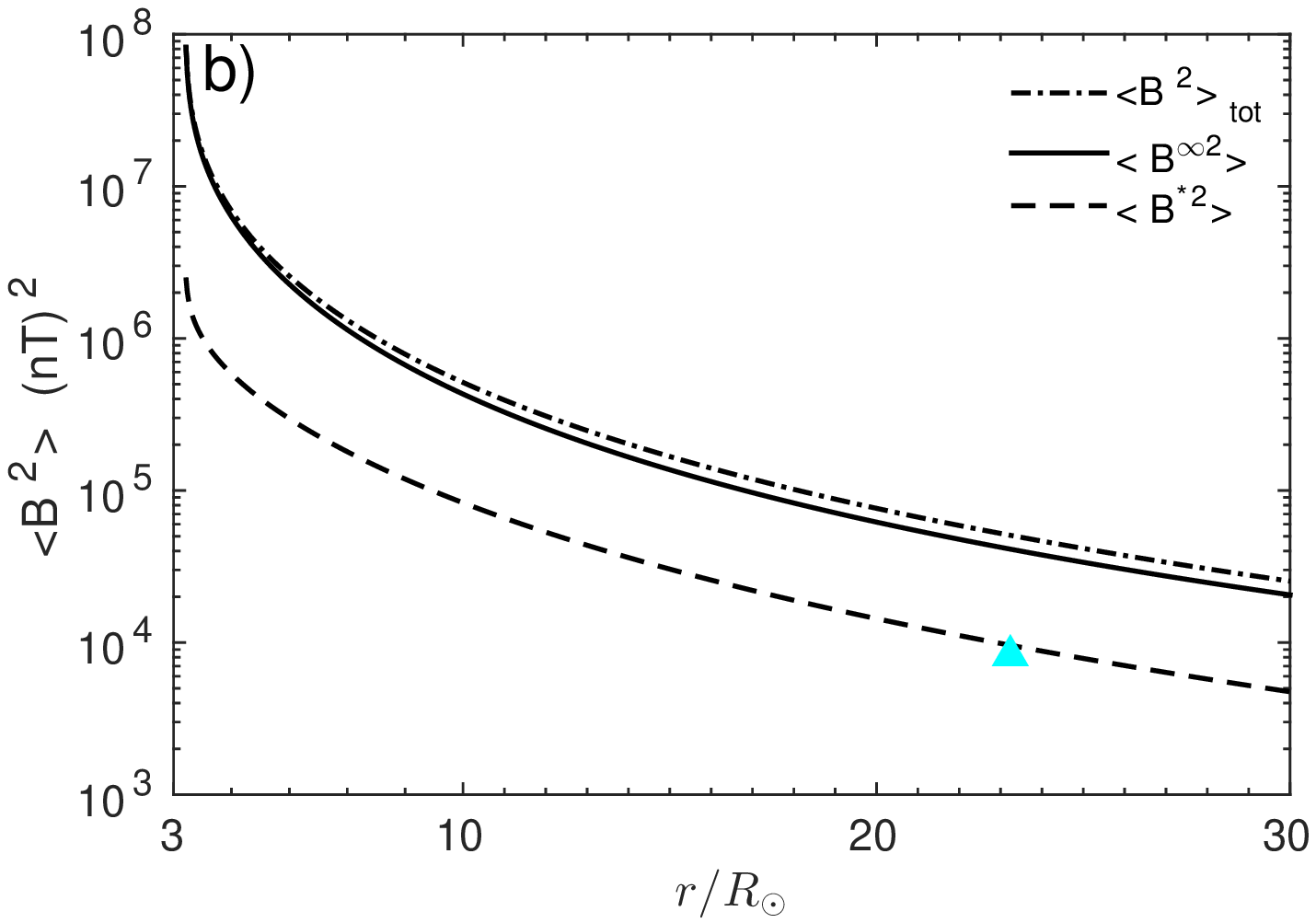}
	\includegraphics[height=0.32\textwidth]{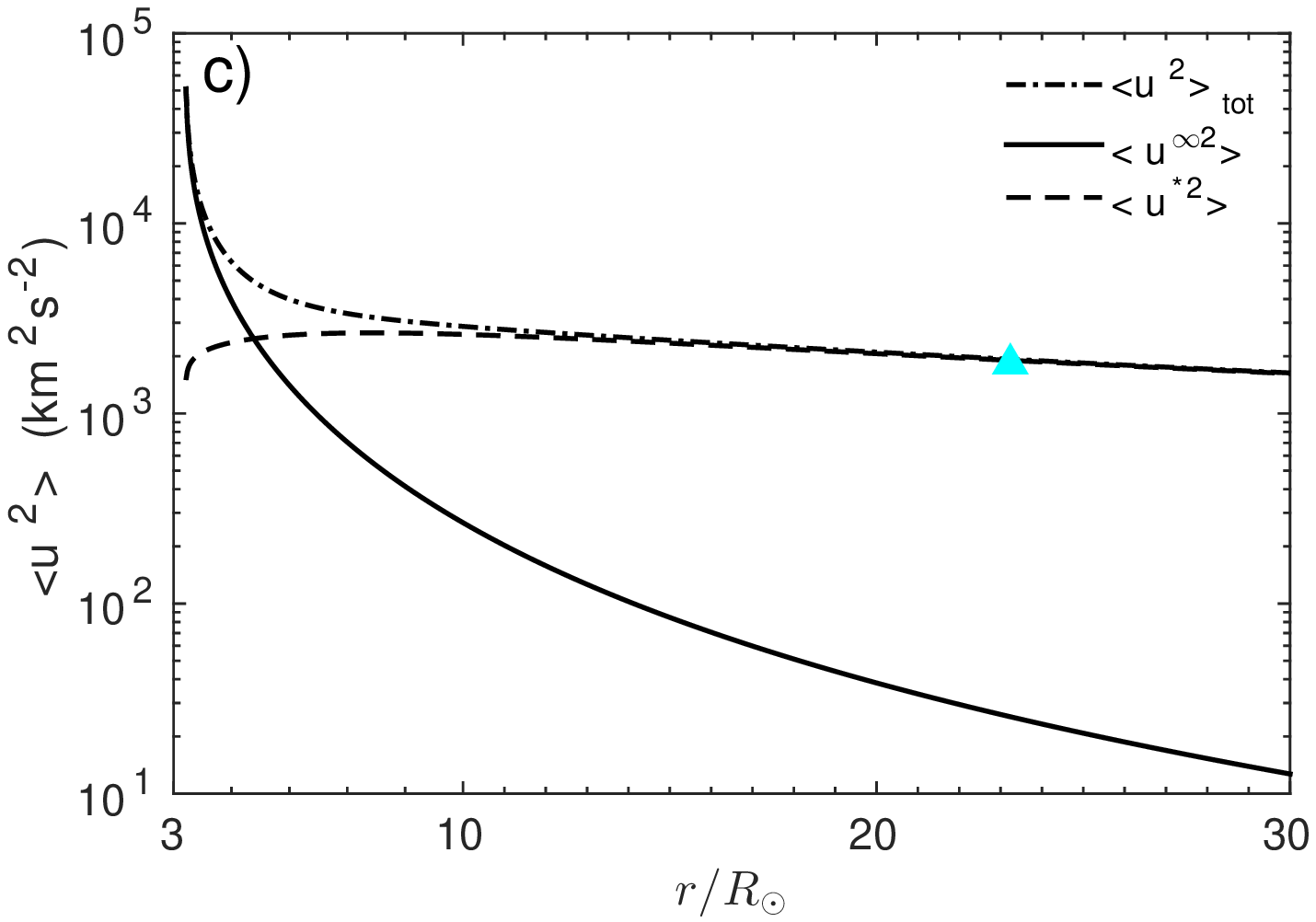}
	\includegraphics[height=0.32\textwidth]{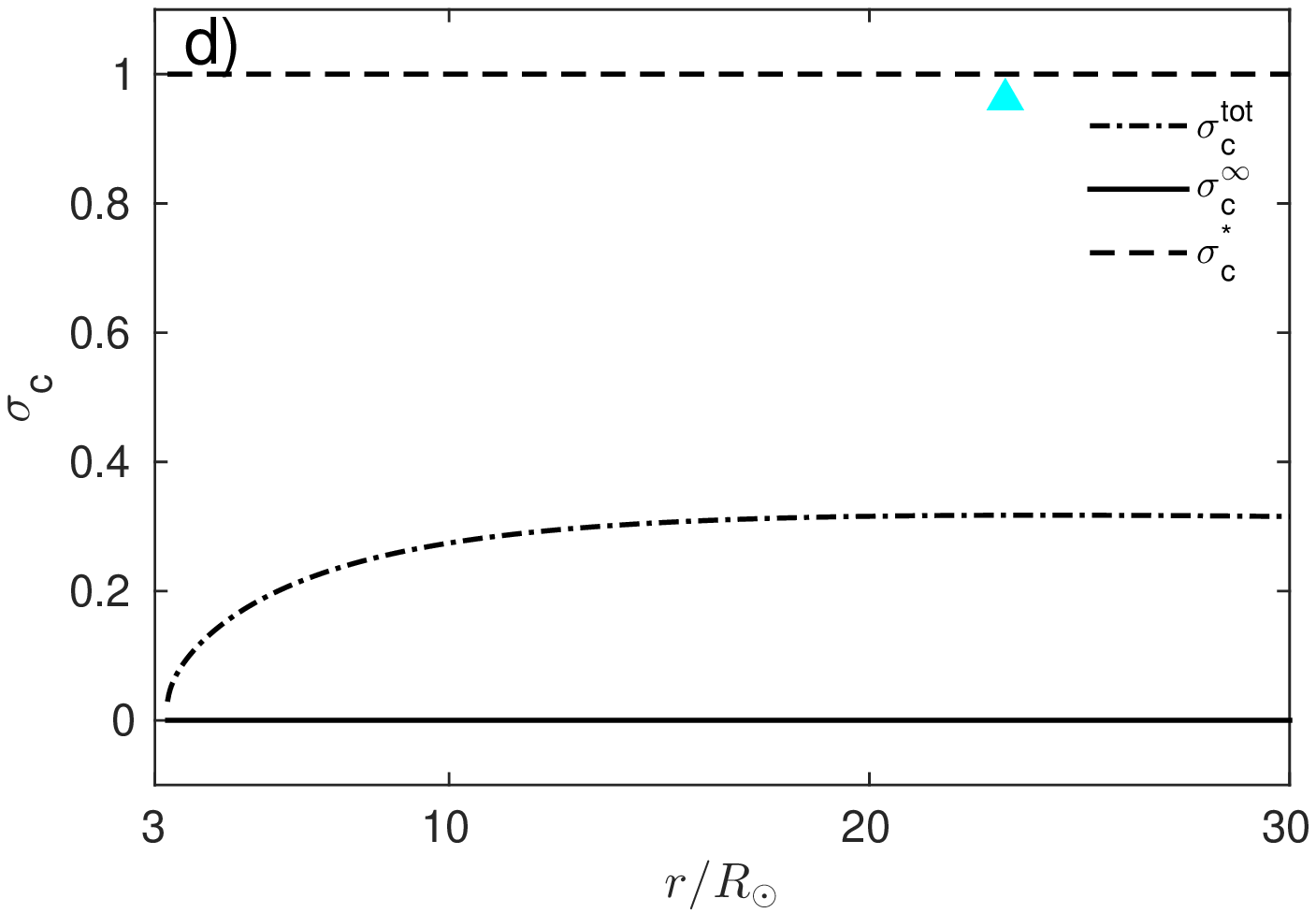}
	\includegraphics[height=0.32\textwidth]{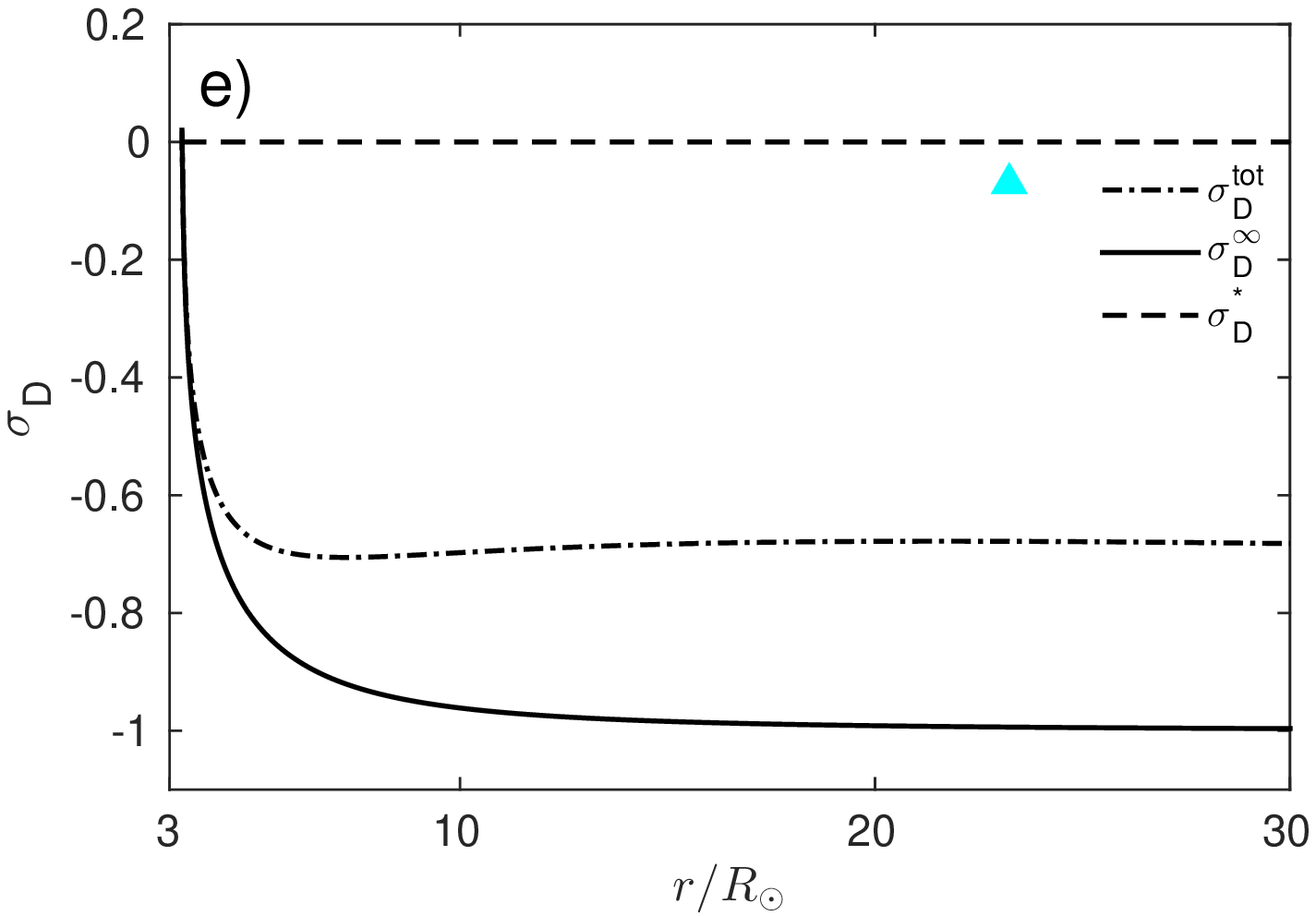}
	\caption{Radial evolution of the total (2D+slab), 2D, and slab turbulence energy from the extended coronal region to the very inner heliosphere. Panels a), b), c), d), and e) describe the outward Els\"asser energy, fluctuating magnetic energy, fluctuating kinetic energy, normalized cross-helicity, and normalized residual energy, respectively. The solid curve represents the 2D component, the dashed curve the slab component, and the dashed-dotted-dashed curve the 2D+slab component. The cyan triangle denotes the corresponding PSP measurement. }
\end{figure}	
\cite{2018ApJ...854...32Z} argued that the magnetic carpet on the photosphere continuously pumps out the 2D structures along with a minority population of Alfv\'en waves (i.e., slab turbulence) above the photosphere. The 2D structures advect through the chromosphere, across the transition region, and into the solar corona. Since advected 2D structures do not reflect at the transition region, there should be no abrupt and significant decrease in the 2D flux at the transition region, as is expected of outward propagating Alfv\'en waves \citep{2018ApJ...854...32Z}. \cite{2018ApJ...854...32Z} argued that the mechanism for heating the solar corona is the same for fast and slow solar wind flow. Recall that the slow solar wind studied in this letter emerges near the equatorial region, and may be  due to the liberation of hot loop material into open field regions by interchange reconnection above the photosphere \citep{2003JGRA..108.1157F}. The slow solar wind in the equatorial region accelerates rapidly within $4$ R$_\odot$, similar to the fast solar wind in open field regions \citep{2020ApJ...901..102A,Telloni2022b}. \cite{2007A&A...476.1341T} found observationally that the solar wind accelerates rapidly within 2 -- 4 R$_\odot$, consistent with Figure 1 (left). The Alfv\'en velocity increases initially to a peak value of $\sim 4 \times 10^2$ kms$^{-1}$, decreases gradually, forming the Alfv\'en surface at $\sim 9.22$ R$_\odot$, and is similar to the in situ PSP-observed Alfv\'en speed (blue full triangle).  

Figure 2 presents the evolution of the basic turbulence quantities. As we see in Figure 2a, the 2D outward Els\"asser energy $\langle z^{\infty + 2} \rangle$ (solid curve) is dissipated rapidly in the extended coronal region, after which it decreases more gradually. By contrast, the minority slab energy in outward propagating modes $\langle z^{* + 2} \rangle$ (dashed curve) increases in the extended solar corona, and decreases beyond the Alfv\'en surface. The initial increase of $\langle z^{* + 2} \rangle$ is due to the presence of the solar wind density gradient term, which acts as source term.
The heating of the solar corona by 2D turbulence is different from that due to counter-propagating Alfv\'en waves \citep{1999ApJ...523L..93M,2010ApJ...708L.116V}, which assumes that a large outwardly propagating Alfv\'enic flux produces reflected Alfv\'en waves that interact nonlinearly to produce 2D modes that dissipates. This physics is in fact incorporated in the slab turbulence equations that comprised NI MHD. The theoretical $\langle z^{* + 2} \rangle$ is similar to PSP measurements, and smaller than the theoretical $\langle z^{\infty + 2} \rangle$. This can be understood by recognizing that the angle between the mean magnetic field and solar wind speed in the interval 18:40 -- 20:40 UT is measured to be $\theta_{UB} \sim 165^\circ$, indicating that PSP measures the slab component, but not the 2D component. The theoretical 2D + slab Els\"asser energy (dashed-dotted-dashed curve) decreases gradually with increasing distance. 

Figure 2b shows that the theoretical 2D fluctuating magnetic energy (solid curve) is larger than the theoretical slab component (dashed curve), and both decrease gradually with increasing distance. The theoretical $\langle B^{* 2} \rangle$ is consistent with the PSP-observed $\langle B^2 \rangle$ for the reason given above. Similarly, the theoretical NI/slab fluctuating kinetic energy $\langle u^{* 2} \rangle$ is consistent with that measured by PSP. The theoretical $\langle u^{* 2} \rangle$ increases initially to a peak value, and then decreases slowly. The theoretical $\langle u^{\infty 2} \rangle$ decreases rapidly, indicating that it is strongly dissipated.

In this model, the slab energy in backward propagating modes $\langle z^{* - 2} \rangle$ is 0, leading to a value of 1 the slab normalized cross-helicity $\sigma_c^*$ (dashed curve in Figure 2c), and is close to the observations. Note that the observed solar wind speed between 3.5 -- 6.3 R$_\odot$ ranges from $96-201$ kms$^{-1}$, and that at 23.4 R$_\odot$ is  219.34 kms$^{-1}$. This is an Alfv\'enic slow solar wind with a high cross-helicity value ($\sigma_c \sim 1$). As shown in Figure 2a, the 2D outward and inward Els\"asser energies are equal, leading to a zero 2D normalized cross-helicity (solid curve in Figure 2d). The total normalized cross-helicity (dashed-dotted-dashed curve) increases gradually until the Alfv\'en surface, and then remains approximately constant. Due to the assumption that the slab turbulent kinetic and magnetic energies are equipartitioned, the slab normalized residual energy (dashed curve) is zero as a function of distance, and is close to the observed value. The 2D normalized residual energy (solid curve) decreases rapidly initially, and then tends to -1, i.e., the dominant 2D component is almost entirely composed of magnetic fluctuations.  

\begin{figure}[h!]
	\centering
	\includegraphics[height=0.32\textwidth]{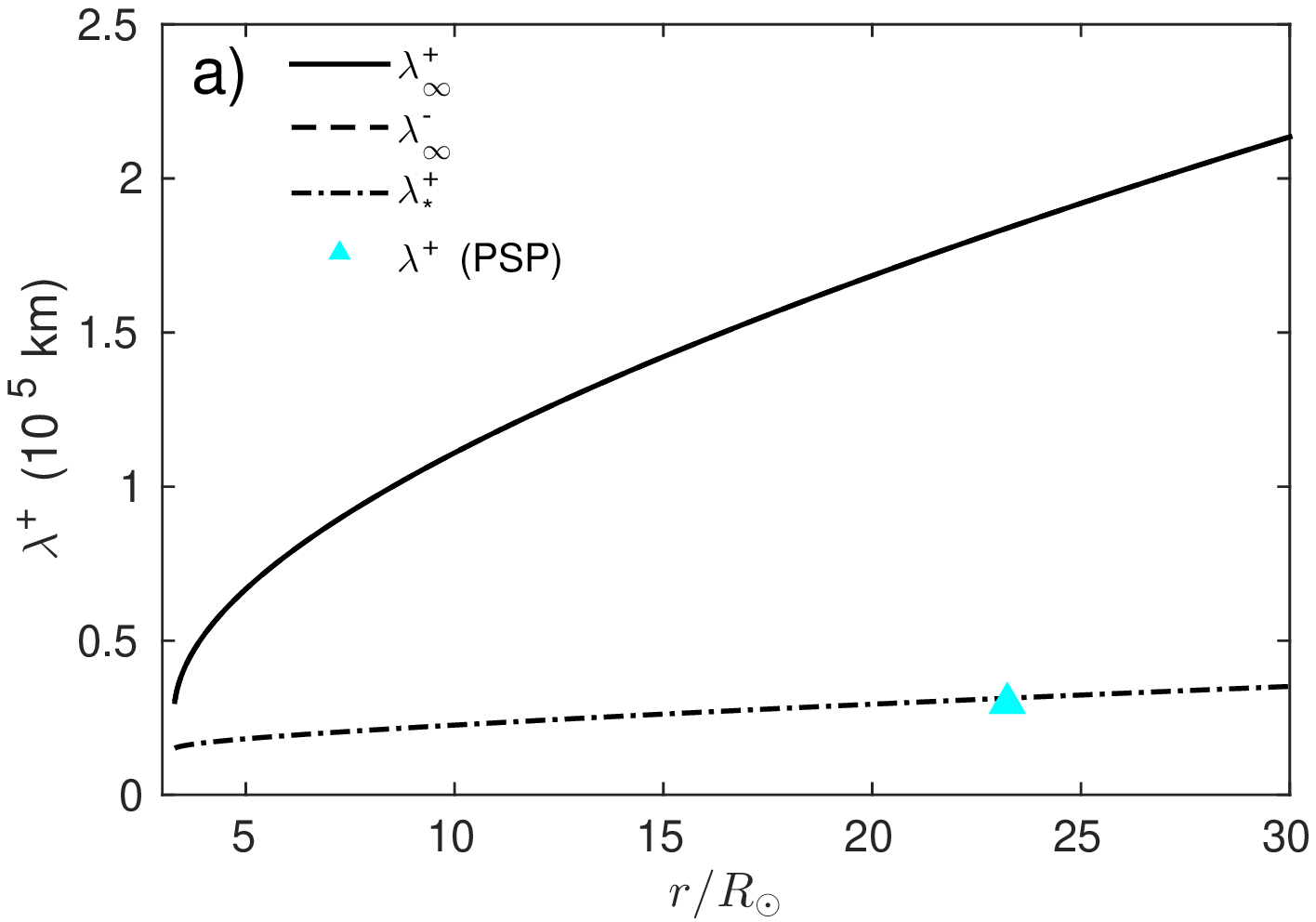}
	\includegraphics[height=0.32\textwidth]{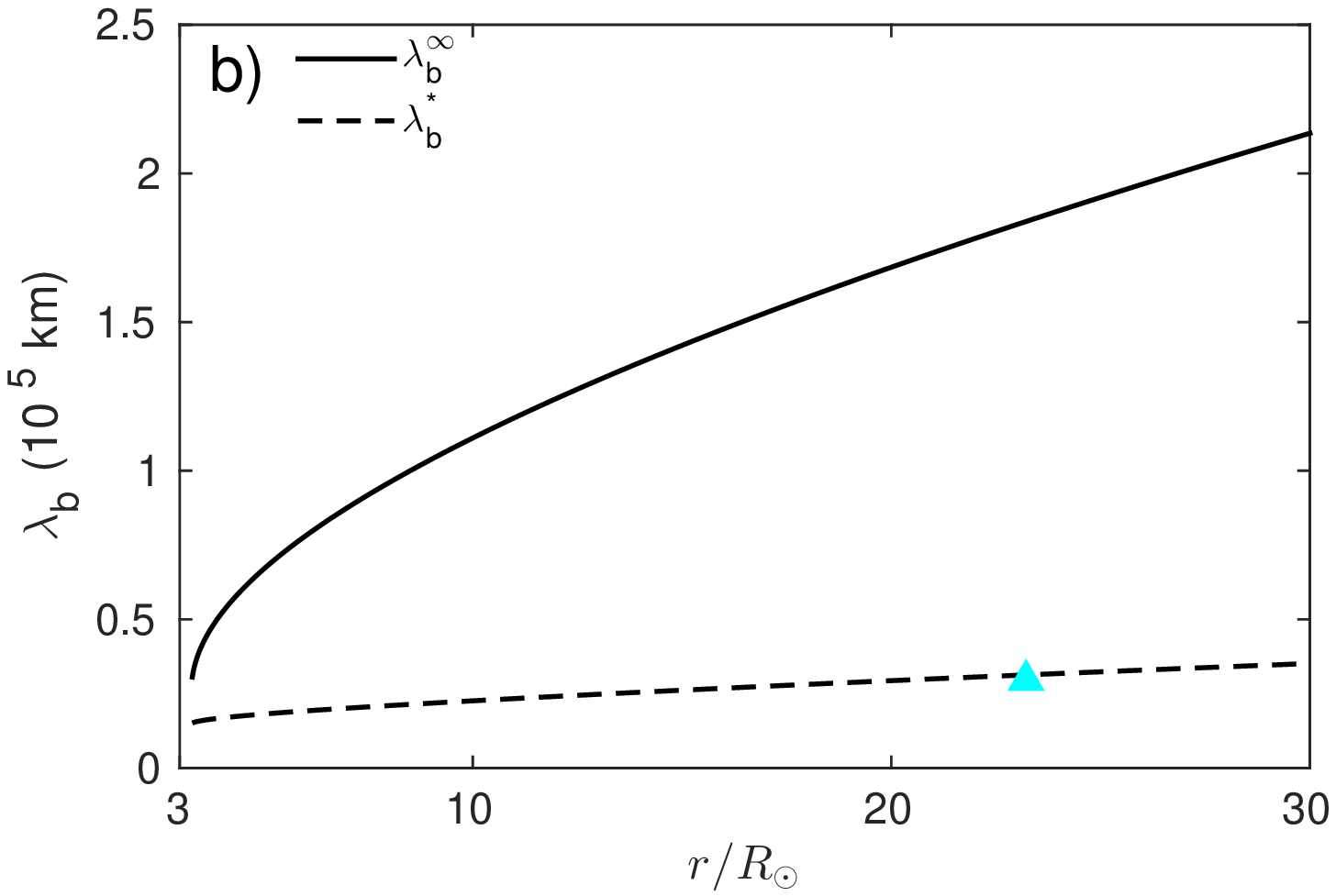}
	\includegraphics[height=0.32\textwidth]{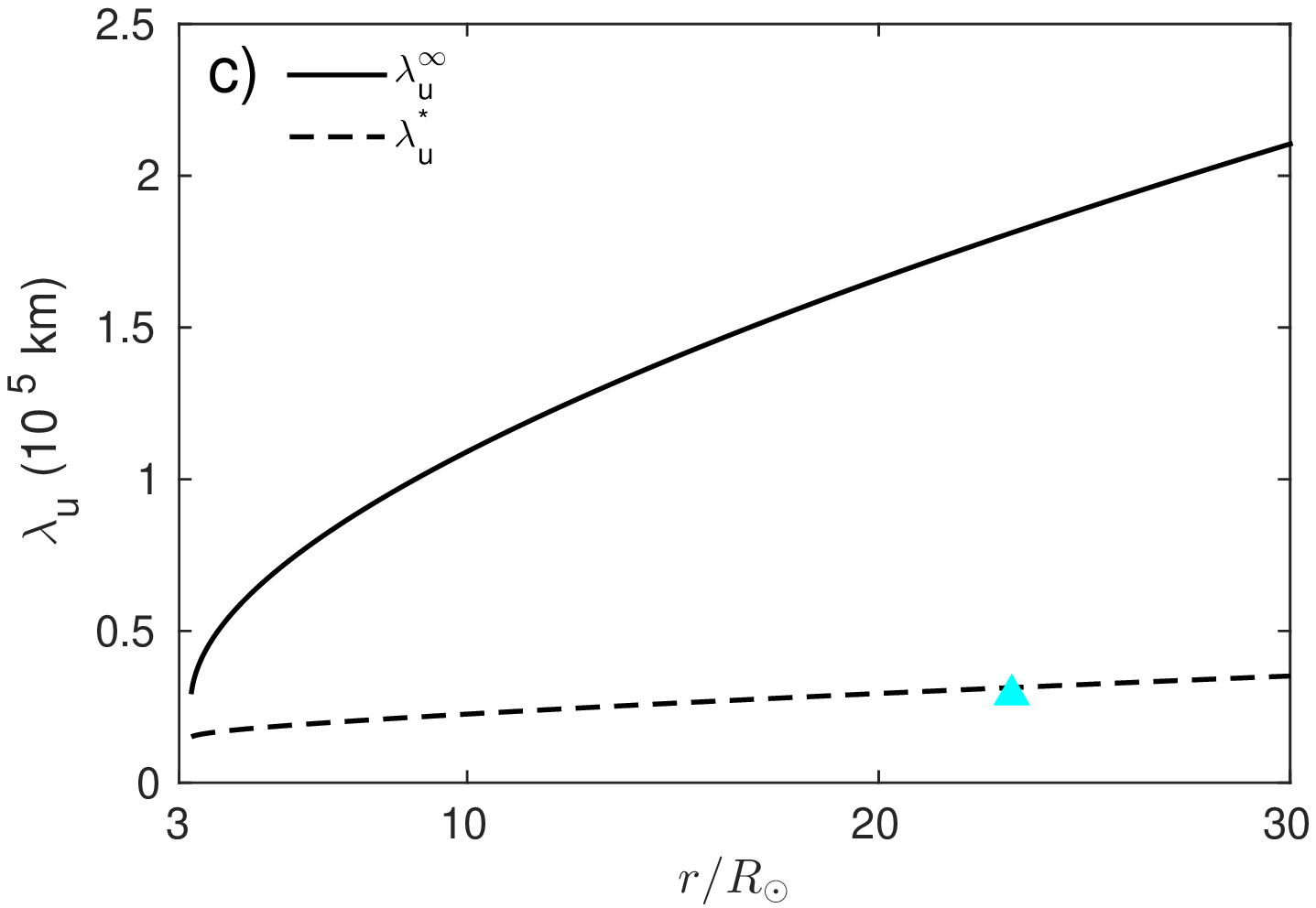}
	\caption{Radial evolution of the 2D (solid curve) and slab (dashed curve) correlation lengths corresponding to the outward Els\"asser energy (a), fluctuating magnetic energy (b), and fluctuating kinetic energy (c). The cyan triangle denotes the PSP measurements. }
\end{figure}
The correlation length is an important parameter because it determines the turbulence heating rate \citep{2021Fluid...6..368A}. Similar to the 2D outward and inward Els\"asser energies, the corresponding 2D correlation lengths are equal (solid black and dashed curves in Figure 3a). The theoretical NI/slab correlation length for the energy in outward propagating modes (dashed-dotted-dashed curve) increases with distance, and is similar to PSP observations. The 2D correlation length of the magnetic field and velocity fluctuations (solid curves in Figures 3b and 3c, respectively) increases much more rapidly than the corresponding slab correlation lengths, the latter of which are in accord with PSP measurements that can measure only the slab fluctuations as discussed above.

\begin{figure}[h!]
	\centering
	\includegraphics[height=0.32\textwidth]{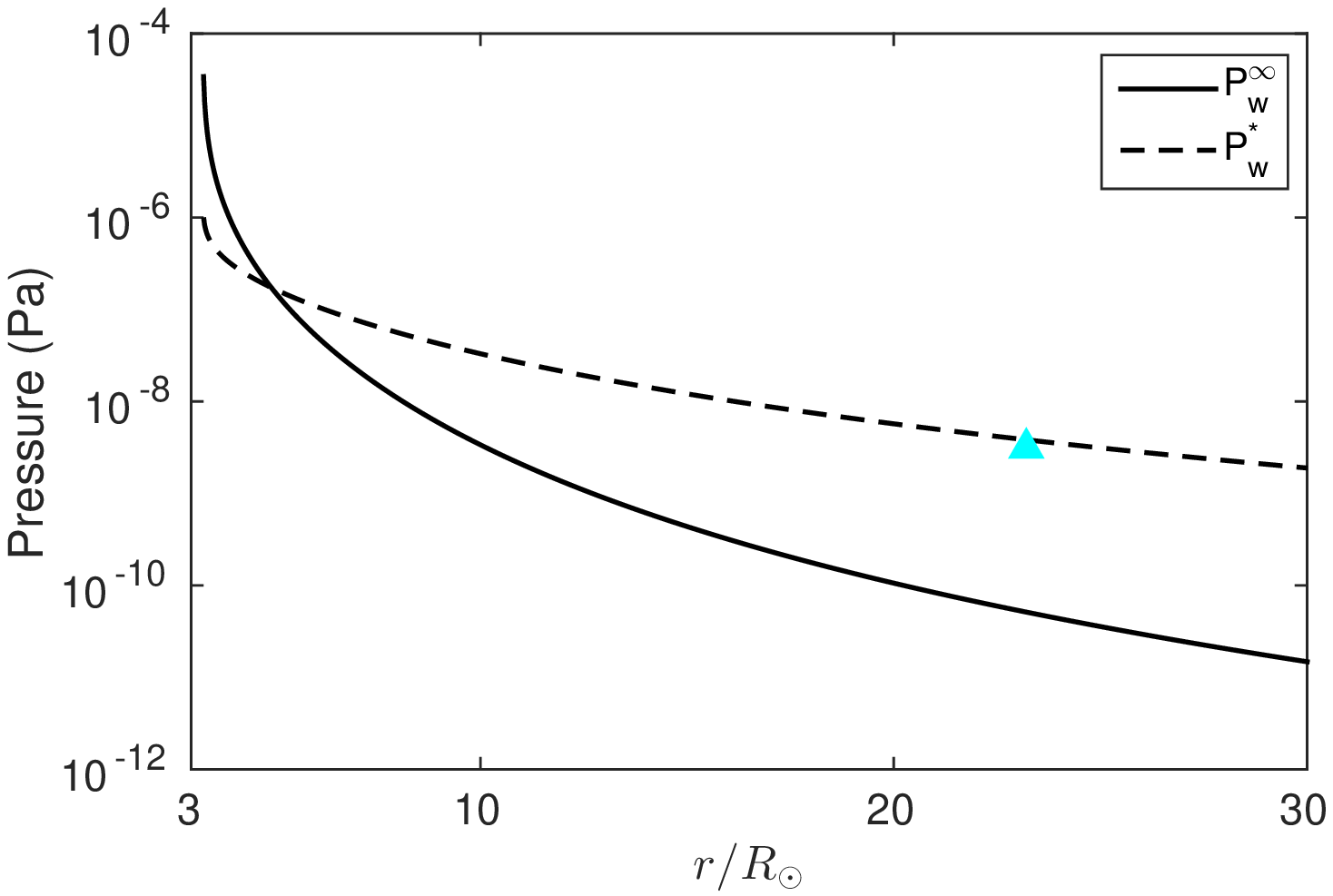}
	\caption{The 2D (solid curve) and slab (dashed curve) turbulence pressure as a function of distance. The cyan triangle is the PSP-observed turbulence pressure. }
\end{figure}
Figure 4 displays the turbulence pressure as a function of heliocentric distance. The solid curve denotes the 2D turbulence pressure and the dashed curve the slab turbulence pressure. Initially, the 2D turbulence pressure is larger than the slab turbulence pressure, however, the prior decreases more rapidly than the latter. The rapid decrease of the 2D turbulence pressure is due to the dominance of the 2D turbulent magnetic energy (see Figure 2e). The theoretical slab turbulence pressure is similar to that observed by PSP at 23.2 R$_\odot$.

From a turbulence perspective, the dissipation of turbulence energy heats the coronal/solar wind plasma. The solar wind proton temperature is assumed to be $7 \times 10^5$ K at 3.3 R$_\odot$, which increases to a peak value of $\sim 1.2 \times 10^6$ K, and then decreases gradually but non-adiabatically in the expanding supersonic solar wind (Figure 5). The theoretical and PSP-observed proton temperature at $\sim 23$ R$_\odot$ are very similar.
\begin{figure}[h!]
	\centering
	\includegraphics[height=0.32\textwidth]{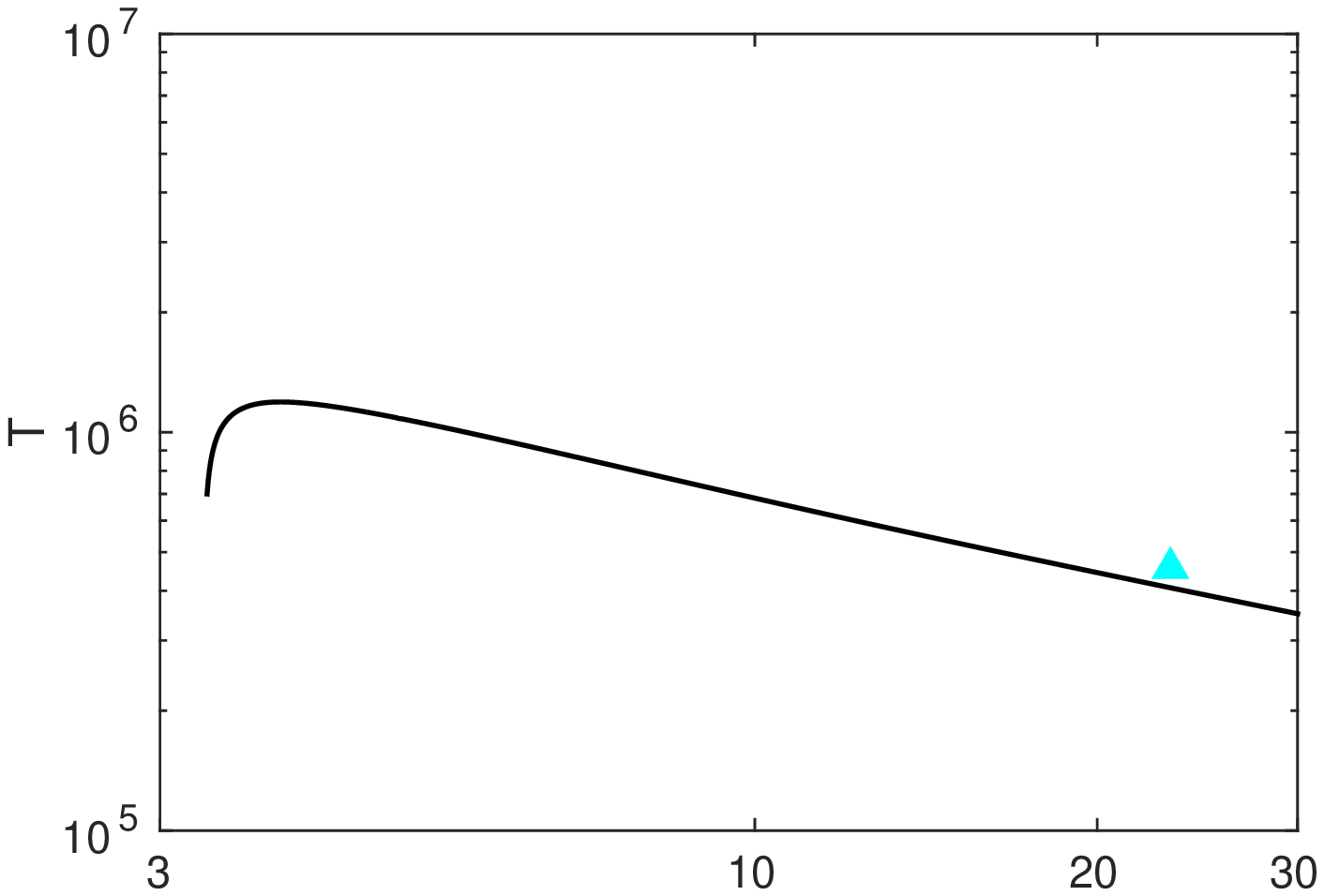}
	\caption{The solar wind proton temperature as a function of distance. The cyan triangle is the PSP-observed proton temperature. }
\end{figure}

\section{Discussion and Conclusions}
\cite{Telloni2022} studied the PSP -- SolO quadrature, combining remote imaging by the SolO Metis coronagraph, and in situ plasma data by PSP. We present a first comparison of the theoretical model and the joint PSP -- SolO observations. We solved a solar wind + NI MHD turbulence model \citep{2020ApJ...901..102A,Telloni2022b} from 3.3 -- 30 R$_\odot$, and compared i) the theoretical solar wind speed and density with the corresponding remote observations from $3.5-6.3$ R$_\odot$ \citep{Telloni2022} and in situ PSP measurements at 23.2 R$_\odot$ or 0.11 au, and ii) the theoretical turbulence energy and correlation length with PSP measurements. We used the PSP magnetometer and SPAN ion plasma data from 18:40 -- 20:40 UT on January 18, 2021 (interval \#1 in \cite{Telloni2022}), and calculated the transverse energy in forward propagating modes, fluctuating magnetic energy, fluctuating kinetic energy, normalized residual energy and cross-helicity, and the corresponding correlation lengths using the method developed by \cite{A2022}. We found that the $\theta_{UB}$, angle between the mean solar wind flow and mean magnetic field for the selected interval is $165^\circ$, indicating that PSP observed primarily the slab turbulence component in this highly field-aligned flow. We found very good agreement between theory and observations. We summarize our findings as follows.     
\begin{enumerate}
	\item The theoretical solar wind speed and density are consistent with those measured by SolO/Metis from $3.5-6.3$ R$_\odot$ and PSP at 23.2 R$_\odot$. The theoretical and observed solar wind speed increases rapidly within 3.3 -- 4 R$_\odot$ ranging from 96 -- 201 kms$^{-1}$, becoming supersonic at $\sim 5.16$ R$_\odot$. Thereafter, the theoretical solar wind speed increases gradually with distance, and is consistent with the PSP speed of 219.34 kms$^{-1}$ measured at 23.2 R$_\odot$. PSP and Metis/SolO measured a slow solar wind stream emerging from the southern coronal hole near the equatorial region \citep{Telloni2022}. The theoretical Alfv\'en velocity increases initially to a peak value $\sim 4 \times 10^2$ kms$^{-1}$, and then decreases gradually to be consistent with that measured by PSP. In this model, the Alfv\'en surface is located at $\sim 9.22$ R$_\odot$.
	
	\item The theoretical 2D outward Els\"asser energy and fluctuating magnetic energy are larger than the corresponding slab components, the latter being close to the corresponding PSP-observed results. Similarly, the theoretical slab fluctuating kinetic energy is also consistent with the PSP-observed kinetic energy at 23.2 R$_\odot$.
	
	\item The theoretical slab normalized cross-helicity is close to the PSP-observed cross-helicity ($\sigma_c = 0.96$), indicating that PSP observed highly Alfv\'enic slow solar wind in the inner heliosphere \citep[e.g.,][]{2019MNRAS.483.4665D}. The theoretical normalized slab residual energy is similar to the PSP-observed residual energy ($\sigma_D\sim -0.07$). 
	
	\item The theoretical 2D correlation lengths corresponding to outward Els\"asser energy, and magnetic field and velocity fluctuations exceed the theoretical slab correlation lengths, and the slab correlation lengths are consistent with those observed by PSP.
	
	\item We derived the two sets of equations in a conservation form, including the super-radial expansion, from the 2D + NI/slab turbulence transport equations that were derived for the unidirectional Alfv\'en waves \citep{2020ApJ...901..102A,Telloni2022b}. Both sets of equations resemble the WKB form in the absence of dissipation term, mixing term, and the turbulence source \citep{2022ApJ...928..176W}. We calculated the theoretical 2D and slab turbulence pressures, and both decrease with increasing distance. The theoretical slab turbulence pressure is similar to that observed by PSP at 23.2 R$_\odot$.  
	
	\item The proton temperature is assumed to be $7 \times 10^5$ K at 3.3 R$_\odot$, increases to a maximum value of $\sim 1.2 \times 10^6$ K, and then decreases gradually with the expanding solar wind. The PSP-measured temperature and the predicted temperature at 23.2 R$_\odot$ are very similar.	
\end{enumerate}
Our theoretical results successfully describe the slow solar wind stream measured by Metis/SolO and PSP from the extended corona to the very inner heliosphere. Future combined studies using combined PSP, SolO, and BeliColombo measurement will be of great value.

\begin{acknowledgements}
	We acknowledge the partial support of a Parker Solar Probe contract SV4-84017, an NSF EPSCoR RII-Track-1 cooperative agreement OIA-1655280, and NASA awards 80NSSC20K1783 and 80NSSC21K1319. The SWEAP Investigation and this study are supported by the PSP mission under NASA contract NNN06AA01C.
\end{acknowledgements}


\end{document}